\documentclass[amsmath,amssymb,aps,twocolumn,pre,floatfix]{revtex4}

\usepackage{mathrsfs}
\usepackage{amsmath}
\usepackage{amssymb}
\usepackage{color}
\usepackage{graphicx}	
\usepackage{bm}
\usepackage{ulem} 
\usepackage{titlesec}
\titleformat{\paragraph}[runin]
{\bfseries\scshape}{\theparagraph}{1em}{}
\usepackage{braket}
\usepackage{multirow}
\usepackage{float}
\usepackage[printwatermark]{xwatermark}
\usepackage{xcolor}

\newcommand{\be}{\begin{equation}}
\newcommand{\ee}{\end{equation}}
\newcommand{\bef}{\begin{figure}}
\newcommand{\eef}{\end{figure}}
\newcommand{\bea}{\begin{eqnarray}}
\newcommand{\eea}{\end{eqnarray}}

\begin{document}
\title{Role of Structural Rigidity and Collective Behaviour in the Molecular Design of Gas Hydrates Anti-Agglomerants}
\author{Fran\c cois Sicard$^{1}$}
\thanks{Corresponding author: \texttt{francois.sicard@free.fr}.}
\author{Alberto Striolo$^{2}$}
\affiliation{$^1$ Department of Physics and Astronomy, University College London, WC1E 6BT London, UK}
\affiliation{$^2$ Department of Chemical Engineering, University College London, WC1E 7JE London, UK}
\begin{abstract}
 Antiagglomerants (AAs) are surface active molecules widely used in the rubber 
 and petroleum industry, among others. In the petroleum industry, it is believed 
 that AAs strongly adsorb to the surface of hydrate particles to prevent the growth 
 of clathrate hydrate within oil pipelines. Small changes in their molecular structures 
 can strongly affect the thermodynamic and kinetic stability of the system as a whole. 
 Here we employ molecular dynamics simulations to study the interplay between the 
 modification of the molecular structure, rigidity and collective effects of AAs 
 designed to prevent hydrate agglomeration in the conditions encountered in rocking 
 cell experiments.
The AAs are surface-active compounds with a complex hydrophilic head and three hydrophobic 
tails whose structural rigidity is enhanced with the attachement of a simple aromatic group. 
 We observe that the aromatic group can positively or negatively affect the performance 
 of the AAs, depending on its location along the hydrophobic tail.
Our approach is based on first quantifying the molecular mechanisms responsible 
for the macroscopic performance. Although the mechanisms at play depend on 
the application, the methodology implemented could be applicable to other 
high-tech industries,  where the agglomeration of small particles must be controlled.

\end{abstract}

\maketitle

Gas hydrates, also known as clathrate hydrates, are ice-like inclusion compounds consisting of 
polyhedral hydrogen-bonded water cages stabilized by guest gas molecules.~\cite{2002-FPE-Koh-Soper,2006-EF-Kelland,2008-CRCPress-Sloan-Koh}
They are formed under high-pressure and low-temperature conditions such as those found in deep oceans 
and pipelines.~\cite{1997-Geology-Brewer-Kirkwood}
Clathrate hydrates are relevant in a variety of scientific and industrial contexts, including climate 
change modeling,~\cite{1996-Paleoceanography-Kaiho-Wada} carbon dioxide sequestration,~\cite{2006-PNAS-Park-Yaghi} hydrocarbon extraction,~\cite{2003-Nature-Sloan} 
hydrogen and natural gas storage,~\cite{2003-Nature-Sloan,2004-Science-Florusse-Sloan,2002-Science-Mao-Zhao} separation and refrigeration technologies,~\cite{2006-ATE-Ogawa-Mori} 
marine biology,~\cite{2000-Naturwissenschaften-Fisher-McMullin} and planetary surface chemistry.~\cite{1974-Science-Milton} 
Of particular interest are the hydrocarbon hydrates that can form blockages in oil and gas pipelines. 
This phenomenon can severely affect the safety of pipeline flow assurance, potentially leading to 
large negative environmental consequences.~\cite{1934-IEC-Hammerschmidt,2002-FPE-Koh-Soper,2008-CRCPress-Sloan-Koh,2019-COCE-Striolo-Walsh}

Among the range of approaches routinely used to manage gas hydrates is the stabilization 
of hydrate particles in hydrocarbon 
dispersions. Specifically designed surfactants, known as anti-agglomerants (AAs), are optimized 
to prevent hydrate plug formation in flow assurance.~\cite{2006-EF-Kelland,2006-CES-Kelland-Chosa,2009-JPSE-Kelland-Andersen,2014-CRCPress-Kelland} AAs are believed to adsorb on 
hydrate particles by their hydrophilic head groups, while the AAs tail groups are soluble in 
the hydrocarbon phase. The hydrate particles are expected to be covered by a film of AAs and oil, 
making them repel each other and remain dispersed.~\cite{2016-PCCP-Phan-Striolo,2017-Langmuir-Bui-Striolo,2018-JPCL-Bui-Striolo,2018-Langmuir-Sicard-Striolo,2020-SciRep-Bui-Striolo} 
Laboratory and field observations alike show that many phenomena determine the performance of AAs. 
For instance, small changes in the molecular structure of the surfactants  
can strongly affect their ability to prevent hydrate plugs 
formation.~\cite{2020-SciRep-Bui-Striolo} Of particular interest in the present work is the fact 
that the stability of water-in-oil emulsions is dependent on the rigidity of the interfacial film, 
which in turn can be determined by the chemical structure and the collective effect of 
the AAs.~\cite{2017-Langmuir-Bui-Striolo,2018-Langmuir-Sicard-Striolo,2020-SciRep-Bui-Striolo}\\

\begin{figure}[t]
\includegraphics[width=0.9 \columnwidth, angle=-0]{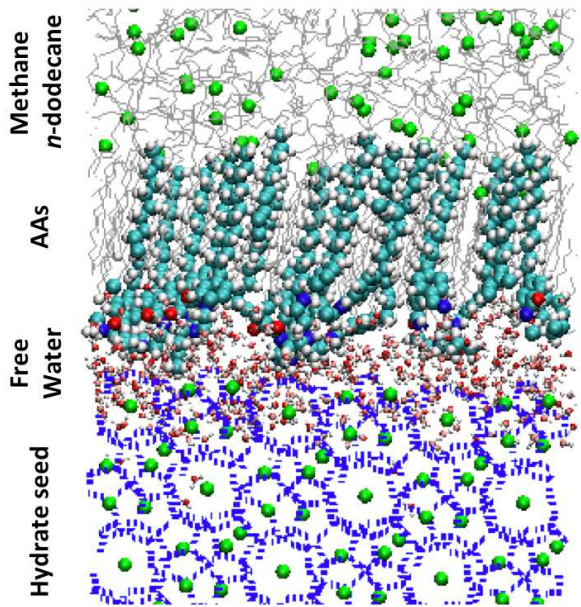}
 \caption{
 Representative simulation snapshot obtained after equilibration for an AA surface density 
 $\approx 0.67~\textrm{molecule/nm}^2$. Green spheres represent methane molecules either in the sII 
 methane hydrate or the hydrocarbon phase. Blue dots lines represent water molecules in 
 the hydrate substrate. Silver lines represent \textit{n}-dodecane molecules, either in 
 the bulk or trapped within the AA layer. Yellow, red, blue, white, and cyan spheres represent 
 chloride ions, oxygen, nitrogen, hydrogen, and carbon atoms in AA molecules, respectively. 
 Only half of the simulation box is shown here for clarity.
}
\label{fig1}
\end{figure}

Here, we employ molecular dynamics (MD) simulations and enhanced sampling techniques 
to study the molecular-level properties along with  thermodynamic and kinetic information of AAs 
specifically designed to prevent hydrate agglomerations in the conditions encountered 
in rocking cell experiments.
The AAs considered are based on a compound, recently developed, which has shown good laboratory performance 
in preventing hydrate formation in light oils.~\cite{2017-Langmuir-Bui-Striolo,2018-Langmuir-Sicard-Striolo,2020-SciRep-Bui-Striolo} 
The molecular structure of the AA tail groups was computationally 
modified to improve the rigidity of the compound, while maintaining the ability of the polar 
head group to adsorb on the hydrate surface.
We analyse the behaviour of the newly designed AA at the hydrate-oil interfaces 
(Figure~\ref{fig1}), in terms of density profiles 
and preferential orientation, and quantify the thermodynamic and kinetic properties associated with 
the transport of the methane molecules across the AA film. 
Details regarding simulation models and algorithms are reported in the Supporting 
Information (SI).
We observe that, at AA densities similar to those considered in rocking cell experiments, 
the attachement of the simple aromatic can positively or negatively affect the performance 
of the AAs depending on its location along the hydrophobic tail. In particular, 
collective and synergistic effects between the newly synthetic AAs and the liquid hydrocarbon 
can yield a free energy (FE) barrier for methane transport through the surfactant layer, 
whose height is significantly higher than the one measured for synthetic AAs routinely used 
in the oil and gas industry.
Based on our interpretation of the simulation results, this suggests that the engineered AAs 
could have much higher performance in preventing the formation of hydrate plugs than existing AAs. 
Experimental verification is required to test our expectations.\\

\begin{figure}[b]
\includegraphics[width=0.85 \columnwidth, angle=-0]{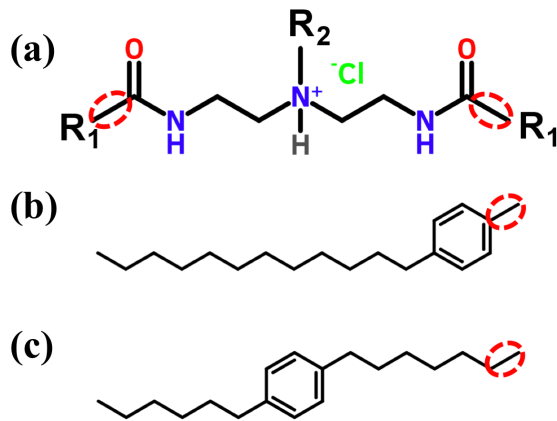}
 \caption{
 (a) Molecular structure of the AAs considered in this work, which contain a headgroup 
 including both amide and tertiary ammonium cation groups, two long hydrophobic tails $R_1$, 
 and one short hydrophobic tail $R_2$. The long $R_1$ tail is composed of a linear 
 hydrocarbon chain of twelve carbon atoms with an aromatic ring (benzene) positioned 
 either at the bottom (b) or in the middle (c) of the continuous chain. The short tail $R_2$ 
 is composed of a linear hydrocarbon chain of four carbon atoms ($n$-butyl). 
 The chain bond connecting the headgroup to the long tails is highlighted in red.
}
\label{fig2}
\end{figure}
\begin{figure*}[t]
\includegraphics[width=0.9 \textwidth, angle=-0]{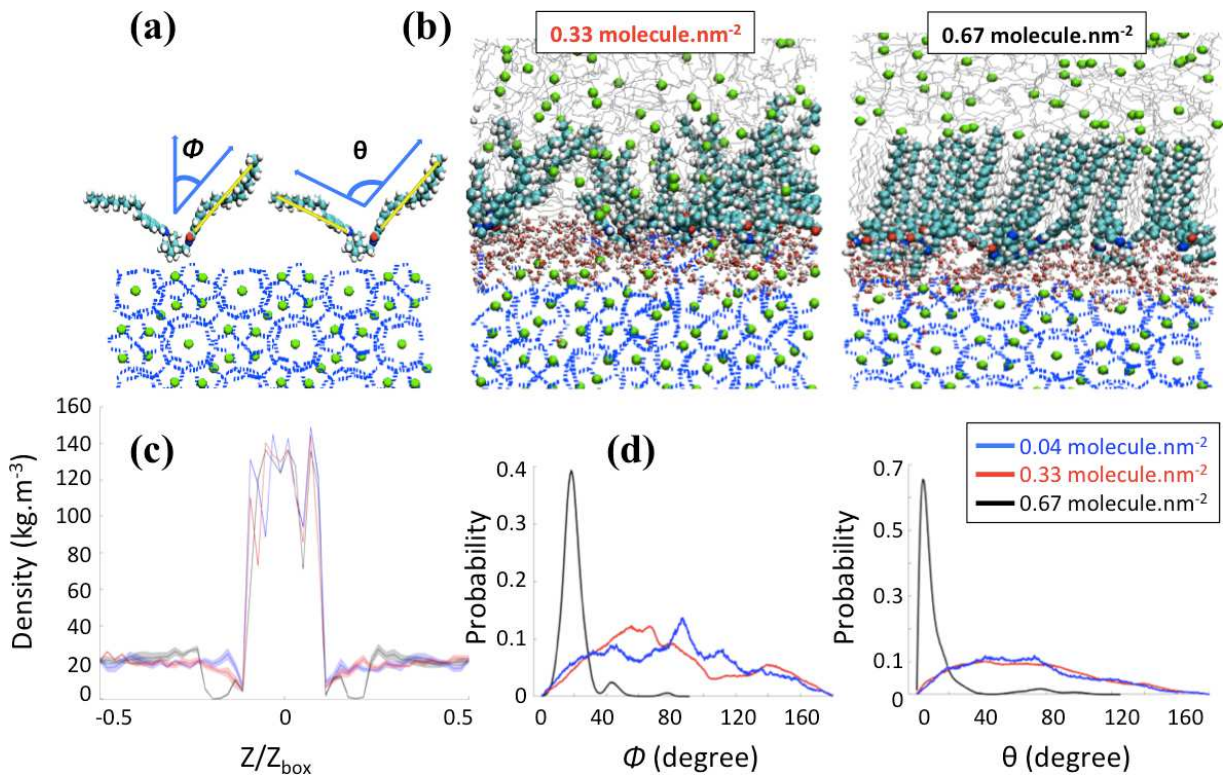}
 \caption{
 (a) Schematic representation of the orientational angle $\Phi$ formed by the vector 
 connecting the first to the last carbons of the hydrocarbon tails and the surface normal 
 ($Z$ direction) and the conformational angle $\theta$ between two long tails of one 
 AA molecule.
 (b) Representative simulation snapshots for the system containing the AA 
 with the aromatic ring  positioned at the bottom of the hydrophobic long tails at two 
 different surface densities. Methane, green spheres; \textit{n}-dodecane, silver lines; 
 water connected by hydrogen bonds, blue lines; chloride ions, yellow spheres. AAs and 
 free water: hydrogen, carbon, oxygen, and nitrogen atoms  are represented by white, cyan, 
 red, and blue spheres, respectively.
 (c) Corresponding density profiles of methane along  the $Z$ direction of the simulation 
 box and (d) probability distributions of orientational ($\Phi$) and 
 conformational ($\theta$) angles at increasing AA surface density.
}
\label{fig3}
\end{figure*}

The two AA molecular structures considered throughout this work are shown in Fig.~\ref{fig2}.
The short tail $R_2$ is a linear hydrocarbon chain of four carbon atoms ($n$-butyl). 
The headgroup including boh amide and tertiary ammonium cation group is known 
for its ability to adsorb on the hydrate surface~\cite{2017-Langmuir-Bui-Striolo,2018-JPCL-Bui-Striolo,2018-Langmuir-Sicard-Striolo,2020-SciRep-Bui-Striolo}. 
 The two long $R_1$ tails are composed of a linear hydrocarbon chain of twelve carbon atoms 
with an aromatic ring (benzene) positioned either at the bottom (Fig.~\ref{fig2}b) 
or in the middle (Fig.~\ref{fig2}c) of the chain. 
The aromatic ring is intended to rigidify the structure of the AA tails 
thus providing significantly stronger mechanical strength and thermal 
stability.~\cite{2015-AOC-Mohanty-Bae}
The simulations were run at temperature and pressure  maintained at $277$ K 
and $20$ MPa, similar to those encountered in laboratory 
experiments~\cite{2017-Langmuir-Bui-Striolo} (see details in the SI). Thus, the conditions 
chosen in this study are well within the gas hydrate stability zone and correspond 
to subcooled systems~\cite{2018-JPCL-Bui-Striolo}.\newline

\textbf{Visual Observation of Simulation Snapshots.} 
We first considered in Fig.~\ref{fig3} the AA structure given in Fig.~\ref{fig2}b, where the aromatic 
ring is positioned at the bottom of the long tail $R_1$. Representative simulation snapshots of 
the AA configurations adsorbed on the hydrate surface are shown in Fig.~\ref{fig3}b at low 
(0.33 molecule.$\textrm{nm}^{-2}$) and high (0.67 molecule.$\textrm{nm}^{-2}$) surface densities. 
The snapshots are taken at the end of our simulations. 
We observe that the headgroups of the AA adsorb on the hydrate surface, as noticed 
previously~\cite{2017-Langmuir-Bui-Striolo,2018-Langmuir-Sicard-Striolo,2018-JPCL-Bui-Striolo,2020-SciRep-Bui-Striolo}. 
The long tails of AAs are instead more likely to extend towards the alkane bulk phase. 
The snapshots shown in Fig.~\ref{fig3}b suggest that different AA densities yield differences in the 
thin-film structure. The increase of the AA surface density yields a transition between disordered 
to ordered orientation of the AA long tails with a thin film within which 
the long tails of AAs and $n$-dodecane align parallel to each other and orient perpendicularly to 
the hydrate surface.
This transition comes with the configurational change of the $n$-dodecane molecules from a 
\textit{Gauche conformation} in the bulk phase to a nearly \textit{all-trans} conformation 
within the AA film.

\begin{figure}[t]
\includegraphics[width=0.99 \columnwidth, angle=-0]{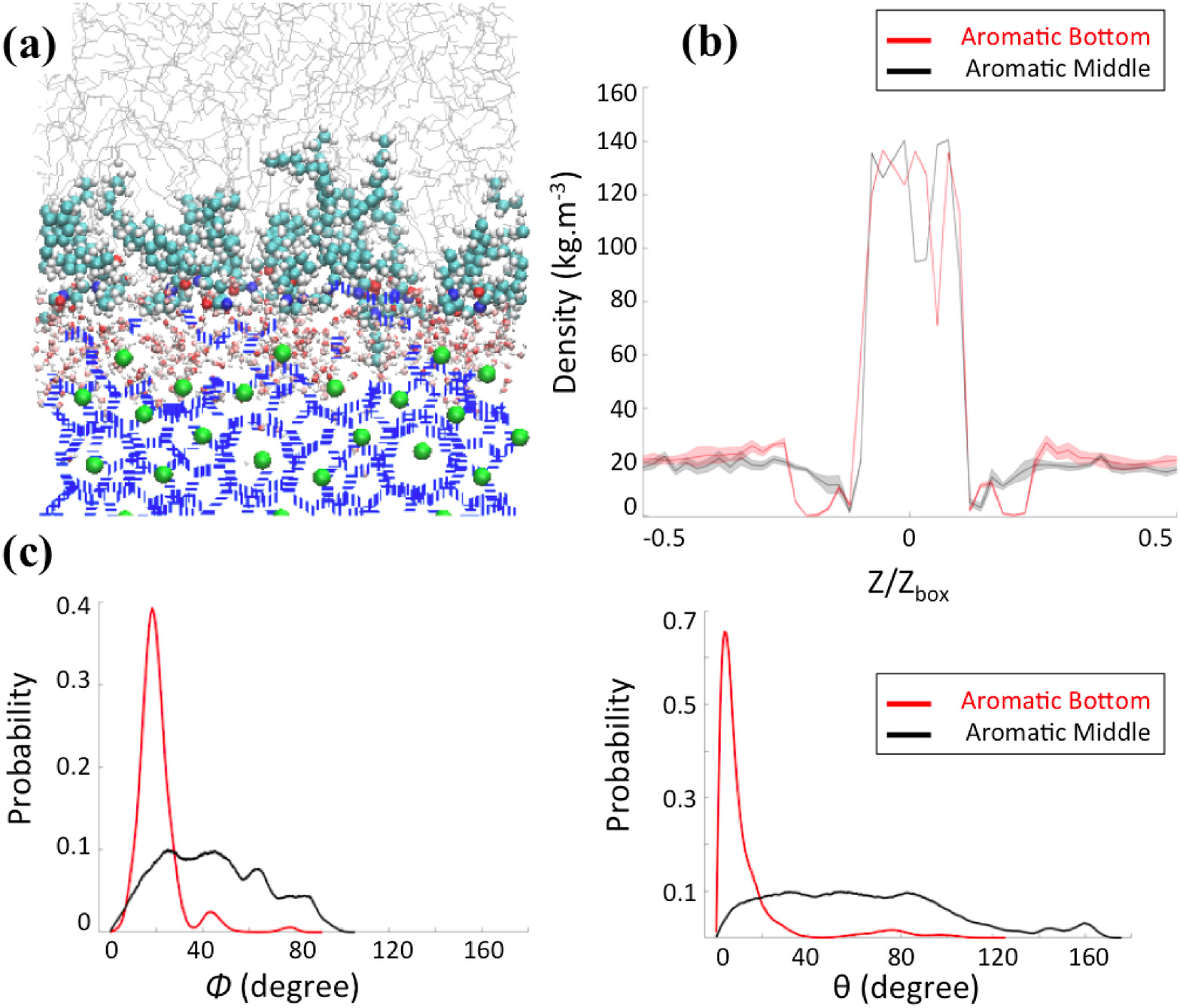}
 \caption{(a) Representative simulation snapshot for the system containing AAs with the aromatic ring 
 positioned in the middle of the hydrophobic long tails at high surface density 
 (0.67 molecule.$\textrm{nm}^{-2}$). The density profiles of the methane along the $Z$ direction (b)
and the probability distributions of orientational and conformational angles (c) at high density 
between systems composed of AAs with the aromatic ring positioned at the bottom (red) or in the middle (black) 
of the hydrophobic long tails are compared.
}
\label{fig4}
\end{figure}

Although changing the position of the aromatic ring from the bottom to the middle in the long tails 
did not seem to affect the ability of the AA to adsorb at the hydrate surface, it significantly affected 
the ordering of the thin-film structure, even at the relatively high surface density 
(0.67 molecule.$\textrm{nm}^{-2}$) studied here, as shown in Fig.~\ref{fig4}a. The lack of ordering 
of the AA film also comes with $n$-dodecane molecules in a \textit{Gauche conformation} either 
in the bulk phase or within the surfactant layer.\newline

\textbf{Density Profiles.} 
To quantify the influence of AAs adsorbed on the hydrate surface on the distribution 
of methane in the system, we calculated the mass density profiles along the $z$-axis 
of the simulation box for increasing AA surface density. 
In the following $Z$ and $Z_{\textrm{box}}$ represent the projection of the Cartesian position of the methane molecules and the size of the simulation box along the $z$-axis, 
respectively.
Figs.~\ref{fig3}c and \ref{fig4}b show a high and constant density between $Z/Z_{\textrm{box}}=-0.1$ 
and $Z/Z_{\textrm{box}}=0.1$, which corresponds to the methane molecules trapped in the hydrate cages. 
For $Z/Z_{\textrm{box}}>0.3$ and $Z/Z_{\textrm{box}}<-0.3$ the results show a uniform density 
representative of the fluid hydrocarbon phase. 
The methane density profile in the thin region between the layer of AA headgroups and the bulk liquid 
hydrocarbon phase shows a pronounced dependence on the AA type and surface density. 
At low AA surface densities, the dendity of methane near the hydrate 
surface for systems containing the aromatic ring either at the bottom (Fig.~\ref{fig3}c) 
or in the middle (Fig.~\ref{fig4}b) of the AA long tails is similar to those found in the bulk. 
When the aromatic ring is positioned at the bottom of the AA long tails (Fig.~\ref{fig3}c), 
the results show a pronounced depletion of methane at the interface ($Z/Z_{\textrm{box}}<0.25$) 
as the AAs surface density increases to 0.67 molecule.$\textrm{nm}^{-2}$, with the density profile 
being nearly $0$. 
Combined with a visual observation of the simulation snapshots, these results suggests that the ordered 
layer successfully expels methane from the interfacial region. This phenomemom is similar to the one 
discussed in previous work for synthetic AA used in the gas and oil 
industry~\cite{2017-Langmuir-Bui-Striolo,2018-Langmuir-Sicard-Striolo}, which can be explained by 
the collective and the synergistic effects between the AAs and the hydrocarbon phase.\newline
 
\textbf{AA Orientation.} 
To quantify the orientation of AAs at the interface, we considered 
the orientational angle, $\phi$, formed between each tail and the direction perpendicular to 
the hydrate surface. We calculated the probability distribution of this angle as well as that of 
the conformational angle, $\theta$, between the two long tails of one AA molecule.
In Figs.~\ref{fig3}d and \ref{fig4}c, we report the probability distribution of the orientational 
and configurational angles for the two AA structures represented in Figs.~\ref{fig2}b and \ref{fig2}c 
at various surface densities. At low surface density, the orientational angle shows a wide probability 
distributions, from $0$ to $90^{\circ}$ and above, irrespectively of the position of the aromatic ring in the 
hydrophobic long tail. 
Similarly, the conformational angle does not show preferential values at low surface coverage 
for either AA. These results suggest that the AAs are rather disordered at these conditions. 
When the AA surface density increases, the results show significant variations. While the results 
obtained for the compound with the aromatic ring positioned in the middle of the long tails 
do not show substantial changes compared to those obtained at low surfae density (Fig.~\ref{fig4}c), 
the results obtained for the compound with the aromatic ring at the bottom show pronounced order. 
As shown in Fig.~\ref{fig3}d, the orientational distribution shows a narrow peak at $\Phi \approx 20^{\circ}$ 
when the surface densiy increases to 0.67 molecule.$\textrm{nm}^{-2}$, suggesting that the AA tails 
become almost perpendicular to the hydrate surface. At the same surface density, the conformational 
distribution shows pronounced peak at $\theta \approx 10^{\circ}$, suggesting that the AAs maintain 
their long tails almost parallel to each other at this conditions.

As the only difference between the two AA structures simulated is the position of the aromatic 
ring in the hydrophobic long tail, the differences highlighted in Fig.~\ref{fig3} and Fig.~\ref{fig4} 
are likely due to collective effects and preferential interactions between the AA long tail 
and the hydrocarbon molecules in the fluid phase due to attractive chain-chain  lateral van der Waals interactions.\newline

\begin{figure*}[t]
\includegraphics[width=0.8 \textwidth, angle=-0]{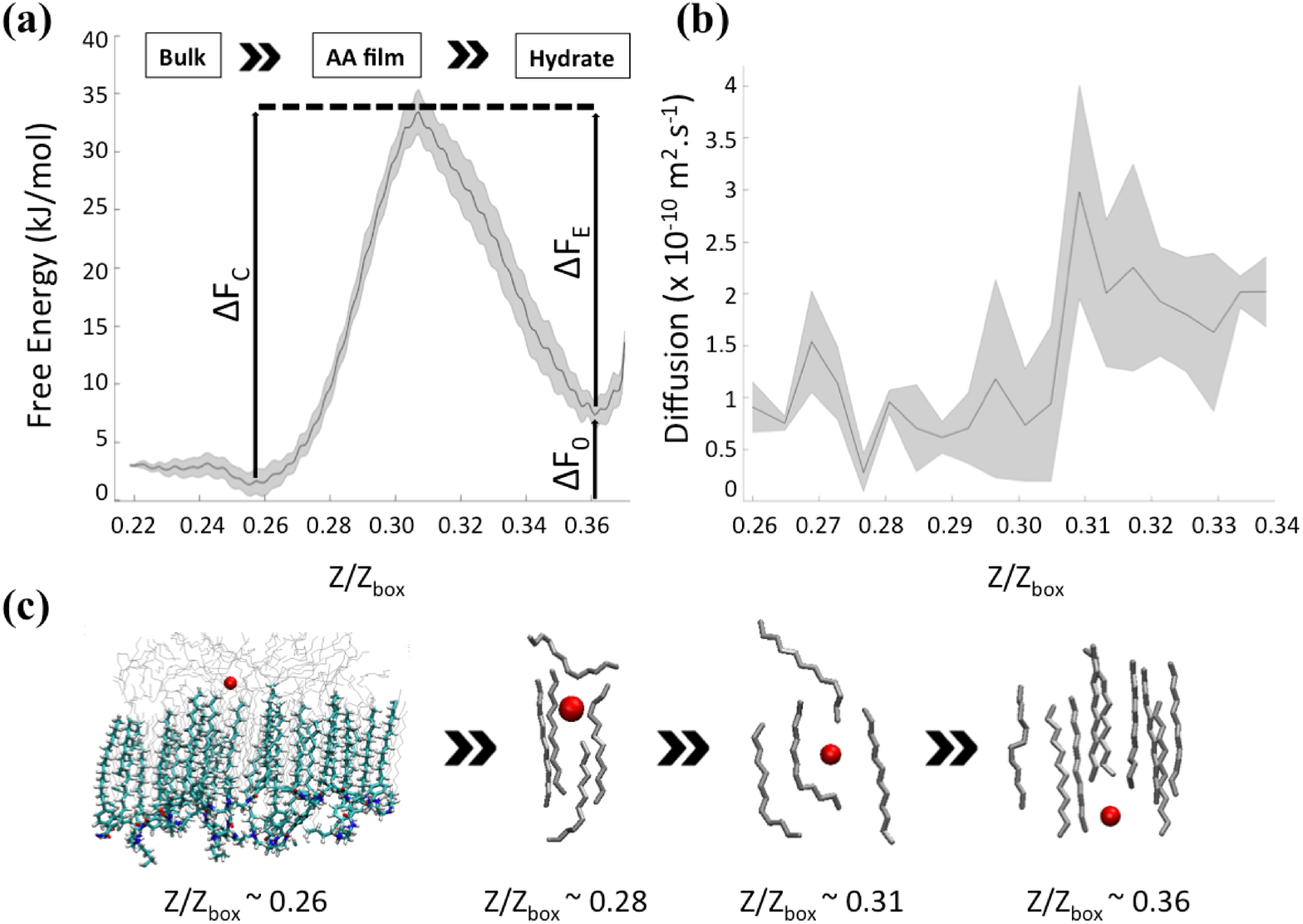}
 \caption{
 (a) FEP associated with the passage of the free methane molecule across 
 the interfacial layer, obtained within the US/ABMD framework and calculated with DHAM 
 and a number of bin of 200. The AA surface density 
 is 0.67 molecule.$\textrm{nm}^{-2}$. The $x$-axis corresponds to the $Z$-Cartesian 
 coordinate of methane expressed in reduced units, $Z/Z_{\textrm{box}}$, 
 with $Z_{\textrm{box}}$ the size of the simulation box along the $Z$ direction. 
 The activation energies associated with methane capture and escape, $\Delta F_C$ 
 and $\Delta F_E$ are $\approx 35~\textrm{kJ/mol}$ and $\approx 26.5~\textrm{kJ/mol}$, 
 respectively. Uncertainties are represented by the shaded area for US data.
 (b) Position-dependent diffusion coefficient calculated from the PACF obtained within 
 the US/ABMD framework. The system shows a diffusion profile with two 
 distinct plateaus at $D\big( Z/Z_{\textrm{box}}< 0.31\big)\approx 0.6\times 
 10^{-10}~\textrm{m}^2.\textrm{s}^{-1}$ and $D\big( Z/Z_{\textrm{box}}< 0.31\big) 
 \approx 2.2\times 10^{-10}~\textrm{m}^2.\textrm{s}^{-1}$ on both sides of the transition 
 state $Z/Z_{\textrm{box}}\approx 0.31$. Uncertainties are represented by the shaded 
  area for US data.
 (c) Sequence of simulation snapshots representing the transport mechanism of methane 
 (red sphere) across the interfacial layer composed of a mixture of AAs and hydrocarbons 
 (silver molecules). The bulk hydrocarbon phase and the sII hydrate are not shown for 
 clarity. The AA layer is only shown in the first snapshot. The methane molecule starts 
 in the bulk hydrocarbon phase, above the AAs layer ($Z/Z_{\textrm{box}}\approx 0.26$). 
 The methane then enters the interfacial layer through two oil molecules. As the methane 
 goes farther across the interfacial layer, the oil molecules bend ($Z/Z_{\textrm{box}} 
 \approx 0.28$), eventually forming a cage surrounding the methane molecule 
 ($Z/Z_{\textrm{box}}\approx 0.31$). As the methane travels farther down, one oil 
 molecule begins pushing the methane molecule. Eventually, methane is driven underneath 
 the AAs layer ($Z/Z_{\textrm{box}}\approx 0.36$).
}
\label{fig5}
\end{figure*}
\textbf{Thermodynamic properties.} 
We quantify the thermodynamic characteristics of the AA layer with the analysis of 
the free energy profile (FEP) associated with the transport of methane molecules. Following previous 
work~\cite{2018-Langmuir-Sicard-Striolo}, we focus our analysis on the diffusion 
of methane through an interfacial region made up of the largest cluster of hydrocarbons 
(characteristic size $\approx 20 \textrm{\AA}$), after the \textit{free} 
methane molecules were expelled from the interfacial region. In this situation, 
the methane molecules interact mainly with hydrocarbon molecules trapped in the AA film. 
In Fig.~\ref{fig5}a, it is shown the FEP obtained within the US/ABMD framework 
employed by Sicard and coworkers,~\cite{2018-Langmuir-Sicard-Striolo,2019-ACSNano-Sicard-Striolo}
and using the dynamic histogram analysis method (DHAM)~\cite{2015-JCTC-Rosta-Hummer} 
plotted along the reduced units, $Z/Z_{\textrm{box}}$ (see details in the SI). 
Uncertainties were determined 
by dividing the data into four equal sections, determining the profiles independently, 
and calculating the standard error. We measured a difference in FE between the global 
($Z/Z_{\textrm{box}}\approx 0.26$) and local ($Z/Z_{\textrm{box}}\approx 0.36$) 
minima $\Delta F_0 \approx 8.5~\textrm{kJ/mol}$. These two basins are 
well separated by activation energies associated with methane capture and escape, 
$\Delta F_C \approx 35~\textrm{kJ/mol}$ and $\Delta F_E \approx 26.5~\textrm{kK/mol}$, 
respectively.
As shown in Fig.~\ref{fig5}c, the free methane molecule is initially in the bulk hydrocarbon phase, 
above the AA layer. When it comes closer to the interface, it is first trapped in a local FE minimum 
($Z/Z_{\textrm{box}} \approx 0.26$). This minimum corresponds to a transition region between 
oil molecules isotropically oriented in the bulk and oil molecules parallel to the AA tails. 
The methane molecule then enters the interfacial film. 
As the methane travels farther across the interfacial layer, an energy barrier arises as the oil molecules 
are displaced from the methane pathway. The transport proceeds until one oil molecule cannot be pushed 
farther down ($0.26\leq Z/Z_{\textrm{box}} \leq 0.31$). Under this conditions, the oil molecule bends, 
eventually forming a cage surrounding the methane molecule ($Z/Z_{\textrm{box}} \approx 0.31$). 
This corresponds to the high-energy transition region in the FEP. Once the methane molecule 
overcomes this transition state, it is pushed down underneath the AA layer. 
The methane molecule then reaches the local minimum corresponding to the water layer between the 
AA layer and the hydrate ($Z/Z_{\textrm{box}} \approx 0.36$).\newline

\textbf{Kinetic properties.}
To complement the thermodynamic analysis, we estimated the position-dependent diffusion profile 
which provides molecular understanding of the transport of solute across three-dimensional 
heterogeneous media.~\cite{2016-JCIM-Lee-Gumbart,2016-JCTC-Gaalswyk-Rowley,2020-arXiv-Sicard-Rosta} 
In this system, the variation of the solute diffusivity can be impacted by variation 
of the frictional environment as the solute moves from bulk hydrocarbon through interface, 
and into the water layer. 
We extended the standard scope of the US framework considering the method 
originated by Berne and co-workers~\cite{1988-JPC-Berne-Straub} and elaborated by 
Hummer,~\cite{2005-NJP-Hummer} where the diffusion coefficient is calculated from the position 
autocorrelation function (PACF) obtained from harmonically restrained simulations
\begin{equation}
D(z_k=\langle z \rangle_k) = \frac{\textrm{var}(z)^2}{\int_0^\infty C_{zz}(t)~dt}~.
\label{Diffeq}
\end{equation}
In Eq.~\ref{Diffeq}, $\langle z \rangle_k$ is the average of the RC in the US window $k$, 
$\textrm{var}(z)=\langle z^2 \rangle - \langle z \rangle^2$ is its variance, and 
$C_{zz}(t)=\langle \delta z(0) \delta z(t) \rangle$ the PACF calculated directly from the time series. 
In Fig.~\ref{fig5}b, it is shown the position-dependent diffusion profile along 
the $Z$ direction of methane across the AAs layer. Uncertainties were determined 
by dividing the data into four equal sections, determining the profiles independently, 
and calculating the standard error.
The system shows a diffusion profile with two distinct plateaus located at positions 
$D\big( Z/Z_{\textrm{box}}< 0.31\big)\approx 0.6\times 10^{-10}~\textrm{m}^2.\textrm{s}^{-1}$ and 
$D\big( Z/Z_{\textrm{box}}< 0.31\big)\approx 2.2\times 10^{-10}~\textrm{m}^2.\textrm{s}^{-1}$ 
on both sides of the transition state $Z/Z_{\textrm{box}}\approx 0.31$.
When methane enters the interfacial layer through two oil molecules, the effective diffusion coefficient 
is similar to the one measured experimentally in bulk hydrocarbons~\cite{1992-SPRINGER-Granick} 
($\approx 5.10^{-11}~\textrm{m}^2.\textrm{s}^{-1}$ at $277$ K and $20$ MPa).
As the methane goes farther across the interfacial layer, the hydrocarbon molecules surrounding it 
start to push it down underneath the AA layer, increasing the effective diffusion coefficient by almost 
an order of magnitude.\\

In conclusion, the extensive simulations discussed above highlight the role 
of the structural rigidity and collective behaviour in the molecular design of new 
gas hydrate AAs. Based on our simulation results, we studied the molecular-level properties 
of a newly synthetic AA specifically designed \textit{in silico} to prevent hydrate plugs formation 
in rocking cell experiments.
We studied the thermodynamic and kinetic characteristics of the system and quantified accurately 
the FEP experienced by one methane molecule travelling across the interfacial film 
along with the position-dependent diffusion coefficient, using a combination of MD simulations 
and enhanced sampling techniques. 
We showed that the FE barrier associated with methane transport across the AAs film 
is due to collective and synergistic effects between the AAs and the liquid 
hydrocarbon trapped in the interfacial layer.\\

Interestingly, the transport properties due to the AAs specifically designed in this work can be compared 
with those associated with synthetic AAs, which show good laboratory performance 
in preventing hydrate formation in light oils~\cite{2017-Langmuir-Bui-Striolo,2018-Langmuir-Sicard-Striolo}. 
In the latter, the calculated FE of activation for methane capture was significantly lower  
($\approx 15~\textrm{kJ/mol}$). It suggests that tuning the rigidity of  AAs at the structural level, 
while preserving their ability to interact with the hydrocarbon molecules, can significantly improve 
their collective performance in preventing hydrate formation.\\

The results presented here follow from the hypothesis that transport trough 
the AAs film determines AAs performance in flow assurance. 
The computational methodology developed could be useful 
for a variety of high-tech technologies in the petroleum, rubber latex, 
ink, and paint and coatings industries, where the agglomeration and 
flocculation of small particles must be controlled~\cite{2008-JCTR-Karlson-Piculell,2017-Langmuir-Striolo-Grady,2019-JCIS-Ardyoni-Eastoe}.
Although the mechanisms by which AAs are effective depends on the application of interest, 
several research and industrial groups focus on the interplay between 
the design and performance of AAs at the microscopic scale~\cite{patentRubber1,patentRubber2,patentHydrate1,patentHydrate2,patentHydrate3}. 
Accounting for the interplay between the structural rigidity and the collective effects 
of new synthetic AAs could allow us to infer their use in practical applications, 
which addresses current industrial needs.

\section*{Acknowledgments}
FS thanks Jhoan Toro-Mendoza, Denes Berta, and Tai Bui for useful discussions.
\textit{Via} our membership of the UKs HEC Materials Chemistry Consortium, which is funded by 
EPSRC (EP/L000202), this work used the ARCHER UK National Supercomputing Service (http://www.archer.ac.uk).
Financial support was graciously provided by the EPSRC to AS under grant number EP/T004282/1.
The project was initiated under the support of EPSRC, via grant number EP/N007123/1.


\pagebreak
\widetext
\begin{center}
\textbf{\large Role of Structural Rigidity and Collective Behaviour in the Molecular Design of Gas Hydrates Anti-Agglomerants} \end{center}

\begin{center}\textbf{\large Supporting Information}
\end{center}
\setcounter{equation}{0}
\setcounter{figure}{0}
\setcounter{table}{0}
\setcounter{page}{1}
\makeatletter
\renewcommand{\theequation}{S\arabic{equation}}
\renewcommand{\thefigure}{S\arabic{figure}}

\section*{Molecular Dynamics (MD) simulations} 
Molecular dynamics (MD) simulations were performed with the GROMACS software 
package, version 5.1.4~\cite{2015-SoftwareX-Abraham-Lindahl} using the TIP4P/Ice water model~\cite{2005-JCP-Abascal-Vega}.
Biased simulations were performed using version 2.3 of the plugin for FE calculation, 
 PLUMED~\cite{2014-CPC-Tribello-Bussi}.
The TIP4P/Ice model has been successfully implemented to study 
hydrate nucleation and growth~\cite{2009-Science-Walsh-Wu,2010-JPCB-Jensen-Sum}  
and to investigate the performance of potential hydrate inhibitors~\cite{2015-PCCP-Alireza-Englezos}. 
This model yields an equilibrium temperature for the formation of gas hydrates at high pressure close to 
experimental values~\cite{2010-JCP-Conde-Vega}. Methane and $n$-dodecane were represented within the united-atom version 
of the TraPPE-UA force field~\cite{1998-JPCB-Martin-Siepmann}. AAs were modeled using the general 
Amber force field (GAFF)~\cite{2004-JCC-Wang-Case}, which is often implemented for modeling organic 
and pharmaceutical molecules containing H, C, N, O, S, P, and halogens. Atomic charges were calculated 
with the AM1-BCC method employed in Antechamber from the Amber 14 suite~\cite{AMBER14}. 
The chloride counterions ($\textrm{Cl}^{-}$) were modeled as charged Lennard-Jones (LJ) spheres with the potential 
parameters taken from Dang~\cite{1994-JCP-Smith-Dang}, without polarizability.
The sII hydrates were considered to be the solid substrate, and they were not allowed to vibrate in this work. 
AAs, chloride counterions, $n$-dodecane, and methane composed the liquid phase. 
Dispersive and electrostatic interactions were modeled by the $12-6$ LJ 
and Coulombic potentials, respectively. The Lorentz-Berthelot mixing rules~\cite{1881-AP-Lorentz,1898-CR-Berthelot} 
were applied to determine the LJ parameters for unlike interactions from the parameters of the pure components. 
The distance cutoff for all non-bonded interactions was set to $1.4$ nm. Long-range corrections 
to the electrostatic interactions were described using the particle mesh Ewald (PME) 
method~\cite{1993-JCP-Darden-Pedersen,1995-JCP-Essmann-Berkowitz,2001-CPL-Kawata-Nagashima} with 
a Fourier grid spacing of $0.12$ nm, a tolerance of $10^{-5}$, and fourth-order interpolation. 
Periodic boundary conditions were applied in three dimensions for all simulations.

\section*{Unbiased MD simulations} 
To construct the initial configurations, we followed the procedure  described in previous work~\cite{2017-Langmuir-Bui-Striolo,2018-Langmuir-Sicard-Striolo}.
The sII methane hydrate was chosen to represent features of the experimental system considered, 
in which a small amount of gases other than methane is present. The underlying assumption 
is that the host gas does not affect the properties of the AAs film, which is the subject 
matter of this investigation.
One unit cell of sII methane hydrates was adapted from the study of Takeuchi et al.~\cite{2013-JCP-Takeuchi-Yasuoka}. 
The sII methane hydrate unit cell was replicated three times 
in the $X$ and $Y$ directions ($5.193$ nm) and two times in the $Z$ direction ($3.462$ nm). 
It was then flanked by a thin liquid water film of approximately 0.5 nm on both sides along 
the Z direction, which represents the quasi-liquid interfacial layer identified in the experiments 
of Aman and coworkers~\cite{2011-PCCP-Aman-Koh}.
The desired number of AA molecules was arranged near both sides of the hydrate substrate. 
The chloride counterions ($\textrm{Cl}^-$) were placed next to the AA headgroups. 
The $n$-dodecane and methane molecules were placed within the remainder of the simulation box.
The time step used in all the simulations was $0.001$ ps, and the list of neighbors was updated every $0.01$ ps 
with the \textit{grid} method and a cutoff radius of $1.4$ nm.\\

The initial configuration was first relaxed using the ``steepest descent minimization" algorithm 
to remove high-energy configurations, which might be related to steric hindrance between AAs 
and the hydrocarbon phase. During this step, the gas hydrate structure remained unaltered.
Subsequently, to minimize the possibility that the initial configuration biased the simulation results, 
an $NVT$ temperature-annealing procedure, as implemented in GROMACS~\cite{2015-SoftwareX-Abraham-Lindahl}, 
was conducted. The algorithm linearly 
decreased the system temperature from $1000$ K to $277$ K in $500$ ps. In these simulations, the hydrate 
substrate and chloride ions were kept fixed in position. To relax the structure of $n$-dodecane and AAs, 
a $NVT$ simulation was conducted at $277$ K for $2$ ns using the Berendsen 
thermostat~\cite{1984-JCP-Berendsen-Haak}, with the sII hydrate structure kept fixed in position.
The equilibration phase was then conducted within the isobaric-isothermal ($NPT$) ensemble under thermodynamic 
conditions favorable for hydrate formation ($T=277$ K and $P=20$ MPa) to equilibrate the fluid density.
During the NPT simulation, all molecules in the system were allowed to move, including water and methane molecules in the hydrate substrate.
The pressure coupling was applied only along the $Z$ direction of the simulation box, which allowed  
the $X$ and $Y$ dimensions to be maintained constant. Temperature and pressure were maintained at $277$ K and $20$ MPa, 
respectively using the Berendsen thermostat and barostat~\cite{1984-JCP-Berendsen-Haak} for $5$ ns. 
This is considered the most efficient algorithm to scale simulation boxes at the beginning 
of a simulation~\cite{2013-Bioinformatics-Pronk-Lindakl}. We then switched to 
the Nose-Hoover thermostat~\cite{1985-JCP-Evans-Holian} and the Parrinello-Rahman 
barostat~\cite{1981-JAP-Parrinello-Rahman} for $100$ ns, which are considered 
 more thermodynamically consistent algorithms~\cite{2013-Bioinformatics-Pronk-Lindakl}. 
This numerical protocol allowed the AAs to assemble and orient to form the interfacial layer 
depicted in Figure~1 in the main text.~\cite{2017-Langmuir-Bui-Striolo,2018-Langmuir-Sicard-Striolo}
The system was then equilibrated for $3$ ns in $NVT$ conditions coupling with the \textit{v-rescale} 
thermostat~\cite{2007-JCP-Bussi-Parrinello} ($T=277$ K, $\tau_T=0.1$ ps). 
To define the position of the methane molecules in the simulation box with respect to the sII hydrate structure, the simulation was continued in $NVT$ conditions holding in place the methane molecule 
enclathrated into the water cages defining the sII hydrate structure.\\

\section*{Biased MD simulations} 
The phenomenon of interest (\textit{i.e.} the transport of  methane 
across the interfacial layer) occurs on time scales that are orders of magnitude longer than 
the accessible time that can be currently simulated with classical MD simulations. 
A variety of methods, referred to as \textit{enhanced sampling techniques}~\cite{2006-CR-Adcock-McCammon,2015-BA-Spiwok-Hosek,2015-BBA-Bernardi-Schulten,2016-Plos-Maximova-Shehu,2017-RP-Pietrucci}, can be implemented 
to overcome this limitation. These methods accelerate rare events and are based on constrained MD. 
In the present work, we used the numerical method employed by Sicard and coworkers,~\cite{2018-Langmuir-Sicard-Striolo,2019-ACSNano-Sicard-Striolo}, which combined the adiabatic biased molecular dynamics 
(ABMD)~\cite{1999-JMB-Paci-Karplus,1999-JCP-Marchi-Ballone,
2011-JCP-Camilloni-Tiana,2016-FD-Sicard-Striolo} and umbrella sampling (US)~\cite{2011-CMS-Kastner} frameworks. 
To design the US windows, we used the projection of the Cartesian position of the methane molecules 
along the $Z$-direction as a reaction coordinate (RC). 
The starting configurations for the US simulations were obtained by pulling adiabatically 
the system along the RC generating $40$ windows. Each US window was subsequently 
run for $2$ ns to allow equilibration, followed by additional $16$ ns of the production run.
To control the accuracy of the sampling with respect to the RC orthogonal degrees of freedom 
(either $X$- or $Y$- direction) we implemented the flat-bottomed potential,~\cite{2004-PNAS-Allen-Roux,
2011-JCTC-Zhang-Voth,2012-JCTC-Zhu-Hummer} as already described in previous work ~\cite{2018-Langmuir-Sicard-Striolo}.\\

Upon completion of the US simulations we obtained the free energy profile (FEP) associated 
with the transport of the methane molecules across the AA film 
using the dynamic histogram analysis method (DHAM).~\cite{2015-JCTC-Rosta-Hummer} 
This unbiasing method, originally derived by Rosta and Hummer, uses a maximum likelihood estimate 
of the Markov State Model (MSM) transition probabilities given the observed transition counts 
during each biased trajectory. To produce an MSM from the enhanced sampling simulations, we first 
discretized the RC into bins, where the number of bins is chosen sufficiently large to give a 
finely discretized coordinate but not so large as to give an under-sampling of transitions 
between bins. Once the bins have been determined, we count the number of observed transitions 
$\big( C_{ij}^k(t)\big)$ between each pair of bins $i$ and $j$ in US simulation $k$ at the chosen 
lagtime $t$, as well as the number of times each bin is occupied $\big( n_i^k = \sum_j C_{ij}^k(t) \big)$ 
during each US simulation $k$. These values then provide the necessary conditional probabilities 
$M_{ij}(t)=P(j,t \vert i,0)$. For biased simulations, where a biasing energy of $u_i^k$ is applied 
to state $i$ during simulation $k$, we computed the unbiased MSM from the biased data 
as given by 
\begin{equation}
M_{ij}(t) = \frac{\sum_k C_{ij}^k(t)}{\sum_k n_i^k~\exp\big( -(u_j^k-u_i^k)/2k_BT\big)}.
\label{DHAMeq}
\end{equation}
Once the MSM has been constructed from simulation data, the equilibrium probabilities 
can be calculated as the eigenvector corresponding to eigenvalue $1$ of the transition 
matrix $M_{ij}$ obtained in Eq.~\ref{DHAMeq}, that is, as its invariant distribution.
To estimate the position-dependent diffusion coefficient associated with the methane transport 
across the interfacial layer, we extended the standard scope of the US framework considering the method 
originated by Berne and co-workers~\cite{1988-JPC-Berne-Straub} and elaborated by 
Hummer,~\cite{2005-NJP-Hummer} where the diffusion coefficient is calculated from the position 
autocorrelation function (PACF) obtained from harmonically restrained simulations
\begin{equation}
D(z_k=\langle z \rangle_k) = \frac{\textrm{var}(z)^2}{\int_0^\infty C_{zz}(t)~dt}~.
\label{Diffeq}
\end{equation}
In Eq.~\ref{Diffeq}, $\langle z \rangle_k$ is the average of the RC in the US window $k$, 
$\textrm{var}(z)=\langle z^2 \rangle - \langle z \rangle^2$ is its variance, and 
$C_{zz}(t)=\langle \delta z(0) \delta z(t) \rangle$ the PACF calculated directly from the time series. 

\normalem
\bibliography{acs}

\end{document}